\documentclass[reprint,superscriptaddress,nofootinbib]{revtex4-2}
\usepackage{graphicx}
\usepackage{amsmath, amssymb, bm, amsbsy}
\usepackage{caption}
\usepackage{siunitx}
\usepackage{lineno}
\usepackage{color}
\usepackage{soul}
\usepackage{mathrsfs}
\usepackage {ulem}
\usepackage{diagbox}
\usepackage{subcaption}
\newcommand{\meqref}[1]{Eq.~(\ref{#1})}
\newcommand{\mpref}[1]{Fig.~\ref{#1}}
\newcommand{\be}{\begin{equation}}
\newcommand{\ee}{\end{equation}}

\usepackage[colorlinks=true, allcolors=blue]{hyperref}

\begin{document}

\title{Emergence of Time Semicrystals in Holographic Driven-Dissipative Systems}

\author{Yu-Qi Lei}
\affiliation{Department of Physics, College of Sciences, Shanghai University, 99 Shangda Road, 200444 Shanghai, China}

\author{Xian-Hui Ge}\email[E-mail: ]{gexh@shu.edu.cn}\thanks{Corresponding author}\affiliation{Department of Physics, College of Sciences, Shanghai University, 99 Shangda Road, 200444 Shanghai, China}

\author{Yu Tian}\email[E-mail: ]{ytian@ucas.ac.cn}\thanks{Corresponding author}\affiliation{School of Physical Sciences, University of Chinese Academy of Sciences, 1 Yanqihu East Road, 101408 Beijing, China}

\author{Shao-Feng Wu}
\affiliation{Department of Physics, College of Sciences, Shanghai University, 99 Shangda Road, 200444 Shanghai, China}

\begin{abstract}
Understanding how temporal order degrades in quantum systems remains a central issue in nonequilibrium physics. Here we study the melting of discrete time crystals in a periodically driven holographic system, where a distinct (discrete) time semicrystal phase emerges with persistent temporal order in disorder, bridging discrete time crystals and fully disordered regimes. This phase exhibits a periodic skeleton, with discrete subharmonic peaks persisting atop a continuous spectrum. We extract a critical scaling behavior across the discrete time crystal to time semicrystal transition. Furthermore, even dynamical transitions between distinct periodic skeletons can be clearly identified with systematic log-periodic corrections to power-law scaling, revealing discrete scale invariance. These findings in holography significantly enrich the platforms for studying nonequilibrium phases of matter.

\end{abstract}

\maketitle

\emph{Introduction}---Nonequilibrium dynamics exhibits a much richer phenomenology than equilibrium physics. The discovery of time crystals, characterized by the spontaneous breaking of continuous or discrete time-translation symmetry~\cite{Wilczek:2012jt,Sacha:2014emr,Else:2016ags,Else:2016yfq}, has stimulated extensive theoretical and experimental research across various quantum many-body systems~\cite{Jiao:2024cvh,Jiao:2024vxx,Greilich_2025,Niemi:2025avm}. Particularly, in Floquet (periodically driven) systems, discrete time crystals (DTCs) are characterized by subharmonic responses. This paradigm has been further expanded by the emergence of higher-order and fractional time crystals~\cite{Pizzi:2021idv,Pizzi:2021hfn,Ye:2021woy,Liu:2024yze}, and time quasicrystals~\cite{Autti:2017jcw,Giergiel:2019bcu,Pizzi:2019gfy,Else:2019alg}. Investigations into these dynamical phases reveal that discrete time crystals can undergo dynamical phase transitions~\cite{Yao:2017dwx,Yao:2018etr,Yousefjani:2024gce}; during this process, strict periodicity is broken, and temporal orders melt, driving the system into disordered (thermalized) dynamics. Recent breakthroughs, such as the multi-stage melting of continuous spacetime crystals~\cite{Liu:2026three} and the order-disorder coexistence in time rondeau crystals (TRCs) which rely on non-periodic yet structured drives~\cite{Moon:2024yxa,Ma:2025qyd}, highlight the intricate breakdown of temporal orders. A critical open question in the intrinsic melting dynamics under standard periodic driving is whether temporal orders melt directly into featureless disorder, or whether these orders have some remnants that are clearly recognizable or even persist as distinct dynamical phases. The possible existence and characterization of such residual phases represent a major frontier in nonequilibrium physics.

Recently, holographic duality (gauge/gravity duality)~\cite{Maldacena:1997re} (for a review, see Ref.~\cite{Zaanen:2015oix}), rooted in foundational macroscopic brane geometries~\cite{Duff:1994an}, provides a very useful platform to explore driven-dissipative dynamics of quantum many-body systems. This framework is natural and fully consistent for studying quantum dissipative dynamics, wherein the black hole horizon acts as a thermal bath and induces dissipation effects on the dual boundary quantum system~\cite{Adams:2012pj}. Currently, holographic duality has been widely applied to the study of nonequilibrium dynamics in many-body systems~\cite{Liu:2018crr}, offering novel insights into nonequilibrium steady states~\cite{Sonner:2012if,Kundu:2019ull,Yang:2024cjk}, far-from-equilibrium holographic transport~\cite{Wondrak:2020tzt}, dynamical phase transitions~\cite{Nakamura:2012ae,Guo:2018mip}, quench dynamics~\cite{Zeng:2022hut,Xia:2023pom,Yang:2025bsw}, quantum turbulence~\cite{Adams:2012pj,Adams:2013vsa,Lan:2016cgl,Zeng:2024rwn}, and pattern formation~\cite{Xia:2024ton}. In particular, recent studies have realized discrete time-translation symmetry breaking within the holographic framework, thereby constructing a dissipative space-time supersolid (STS)~\cite{Yang:2023dvk}. In this context, the dynamical response of the bulk matter field to periodic driving constitutes the holographic dual of synchronization phenomena and discrete time crystalline dynamics on the boundary. Despite this progress, only the simplest second-order discrete time crystalline behavior has been observed in holography, mainly due to the fact that the holographic system in~\cite{Yang:2023dvk} is inhomogeneous in two spatial dimensions and thus too complicated and numerically expensive to be investigated thoroughly. As well, it is extremely difficult to quantitatively study the critical behavior in the nonequilibrium phase transition between the STS and other phases.

In this Letter, we introduce a minimal holographic system with spontaneous $Z_2$ symmetry breaking and study its nonequilibrium dynamics under periodic driving, revealing a rich nonequilibrium phase diagram. Remarkably, we uncover the emergence of novel nonequilibrium phases with residual temporal orders during the melting of DTCs into fully disordered dynamics, termed (discrete) time semicrystals (TSCs), alongside the first realization of holographic higher-order DTCs. We focus our investigation on the formation mechanism and quantitative characterization of the TSC phase, the power-spectrum criterion of which is identified as the coexistence of a periodic skeleton (discrete subharmonic peaks) representing the residual temporal order and a continuous background spectrum. It can be clearly demonstrated that the TSC is not a structureless smooth crossover but a distinct nonequilibrium dynamical phase. Actually, by introducing the Lyapunov exponent as a dynamical order parameter, we track the onset of temporal disorder and extract the critical scaling relation during the melting process from the DTC into the TSC. Furthermore, deep within the TSC phase, tracking the subharmonic peaks in the power spectrum reveals that the dynamical reorganization between different periodic skeletons exhibits critical scaling behaviors characterized by log-periodic corrections, reflecting discrete scale invariance (DSI). Building upon this, we deeply dissect the melting of temporal order as it transitions toward fully disordered dynamics, thereby providing crucial theoretical insights into the fundamental physical question of whether residual periodic order persists within chaos.

\emph{Setup}---We consider a minimal holographic model to study dissipative nonequilibrium dynamics, described by the Einstein-scalar action~\cite{Faulkner:2010gj}
\be
S=\int d^4x\,\sqrt{-g}\left[R+\frac{6}{L^2}-(\nabla\phi)^2 - V(\phi)\right],
\ee
where we set the AdS radius $L=1$, and $R$ is the Ricci scalar. The real scalar field $\phi$ has a nonlinear potential $V(\phi)=m^{2}\phi^{2}+\frac{\phi^{4}}{6}$, where $m$ denotes the effective mass. We work in the probe limit, assuming that the backreaction of the scalar field on the background spacetime is negligible, which is valid outside the extreme low-temperature regime~\cite{HHH}. This setup naturally corresponds to a boundary quantum system coupled to an infinitely large thermal bath, providing an intrinsic mechanism for dissipation.

The dynamics of the bulk scalar field are governed by the covariant Klein-Gordon equation
\be
\nabla^2\phi-\frac{1}{2}\frac{\partial V(\phi)}{\partial \phi}=0.
\ee
The background spacetime is fixed to be a planar Schwarzschild-AdS black hole. In ingoing Eddington-Finkelstein coordinates, the metric reads
\be
d s^{2}=\frac{1}{z^{2}}\left(-h(z)d t^{2} - 2dtdz + d x^{2}+d y^{2}\right),
\ee
with the blackening factor $h(z)=1-z^3$. The AdS boundary is located at $z=0$, and the black hole horizon is at $z_h=1$, yielding a Hawking temperature $T_H = \frac{3}{4\pi}$.

We fix the effective mass $m^2=-2$ and adopt the \textit{alternative quantization}. Near the AdS boundary ($z \rightarrow 0$), the scalar field admits the asymptotic expansion
\be
\phi=z\left(\mathcal{A} + z \mathcal{B} +\cdots \right),\label{Sasy}
\ee
where $\mathcal{A}$ corresponds to the condensate response $\langle \mathcal{O} \rangle$. To induce spontaneous $Z_2$ symmetry breaking, a double-trace deformation is introduced into the dual field theory. This modifies the dual field theory action through a scalar operator $\mathcal{O}$
\be
S \rightarrow S - \int d^3x \, \varkappa_d \, \mathcal{O}^2,
\ee
where $\varkappa_d=2(3-2\Delta_-)\varkappa$ is the coupling parameter, and $\varkappa$ is a rescaled coupling parameter used for convenience. This deformation imposes a mixed boundary condition on the bulk scalar field near the AdS boundary, yielding $\mathcal{B}=\varkappa\,\mathcal{A}$ for the asymptotic behavior in \meqref{Sasy}.

Defining the effective parameter $\bar{\varkappa}=\varkappa / z_h$, the $Z_2$ symmetry is spontaneously broken for $\bar{\varkappa} \lesssim -0.386$ in the equilibrium. Here we set $\bar{\varkappa}=-2$ to maintain spontaneous $Z_2$ symmetry breaking. To drive the system out of equilibrium, we introduce a periodic drive $F_d(t)=F_0 \sin(\omega_d t)$ into the boundary condition at $z=0$. We denote the drive frequency as $f_d=\omega_d/(2\pi)$ and the corresponding period as $T_d=2\pi/\omega_d=1/f_d$. For the rescaled field $\psi=\phi/z$, the driving boundary condition follows~\cite{Lan:2016cgl}
\be
\partial_t \psi\big|_{z=0} = \left( \partial_z \psi - \bar{\varkappa} \psi + F_d(t) \right)\bigg|_{z=0}.
\label{eq:drive_bc}
\ee
From the numerical solutions of the bulk dynamics, we extract the boundary order parameter as $X(t) \equiv \psi(t, z=0)$. All simulations are performed at fixed Hawking temperature $T_H=3/(4\pi)$, fixed $\bar{\varkappa}=-2$, and fixed drive amplitude $F_0=4$, with only the drive angular frequency $\omega_d$ tuned to map the dynamical phase diagram. Detailed partial differential equations and numerical implementations are provided in the supplementary material (SM).

\emph{Numerics and Time Semicrystal}---We explore the dynamical signatures of the ``time semicrystal'' — a distinct nonequilibrium phase emergent in periodically driven dissipative many-body systems. By solving the holographic equation of motion for a bulk scalar field $\psi(t,z)$, we extract the boundary condensate $X(t) \equiv \psi(t, z=0)$ as our primary dynamical observable. The time semicrystal phase exhibits pronounced characteristics that fundamentally distinguish it from both discrete time crystals and full chaos. Mathematically, this phase is strictly characterized by the power spectral density (PSD) $P(f)$ of the boundary response $X(t)$, which takes the form
\be
P(f) = \lim_{T \to \infty} \frac{1}{T} \left| \int_0^T dt \, X(t) e^{-i 2\pi f t} \right|^2.
\ee
In the time semicrystal phase, the power spectral density can be decomposed into a hybrid structure
\be 
P(f) = \sum_{k} A_k \delta(f - f_k) + P_{\text{continuous}}(f).
\ee
In finite-time numerical simulations, these delta functions manifest as sharp subharmonic peaks with finite resolution. The discrete subharmonic peaks at $f_k = \frac{q}{n}f_d$ (with coprime integers $q$ and $n$) reveal a periodic skeleton, while the continuous term $P_{\text{continuous}}(f)$ captures the intrinsically generated disorder, distinguishing it from the externally injected randomness of time rondeau crystals~\cite{Moon:2024yxa,Ma:2025qyd}. Characterized by the fundamental locked frequency $f_d/n$, we label such a state as an \textbf{``$n$-time semicrystal''}. Physically, this indicates the coexistence of temporal order and chaos. The periodic skeleton acts as a remnant of time crystal order with asymptotic periodicity~\cite{PROVATAS1991295}. The defining feature of the time semicrystal phase is the interplay between partial discrete time-translation symmetry breaking and disordered dynamics.

\begin{figure}[htp!]
    \centering
    \includegraphics[width=0.95\linewidth]{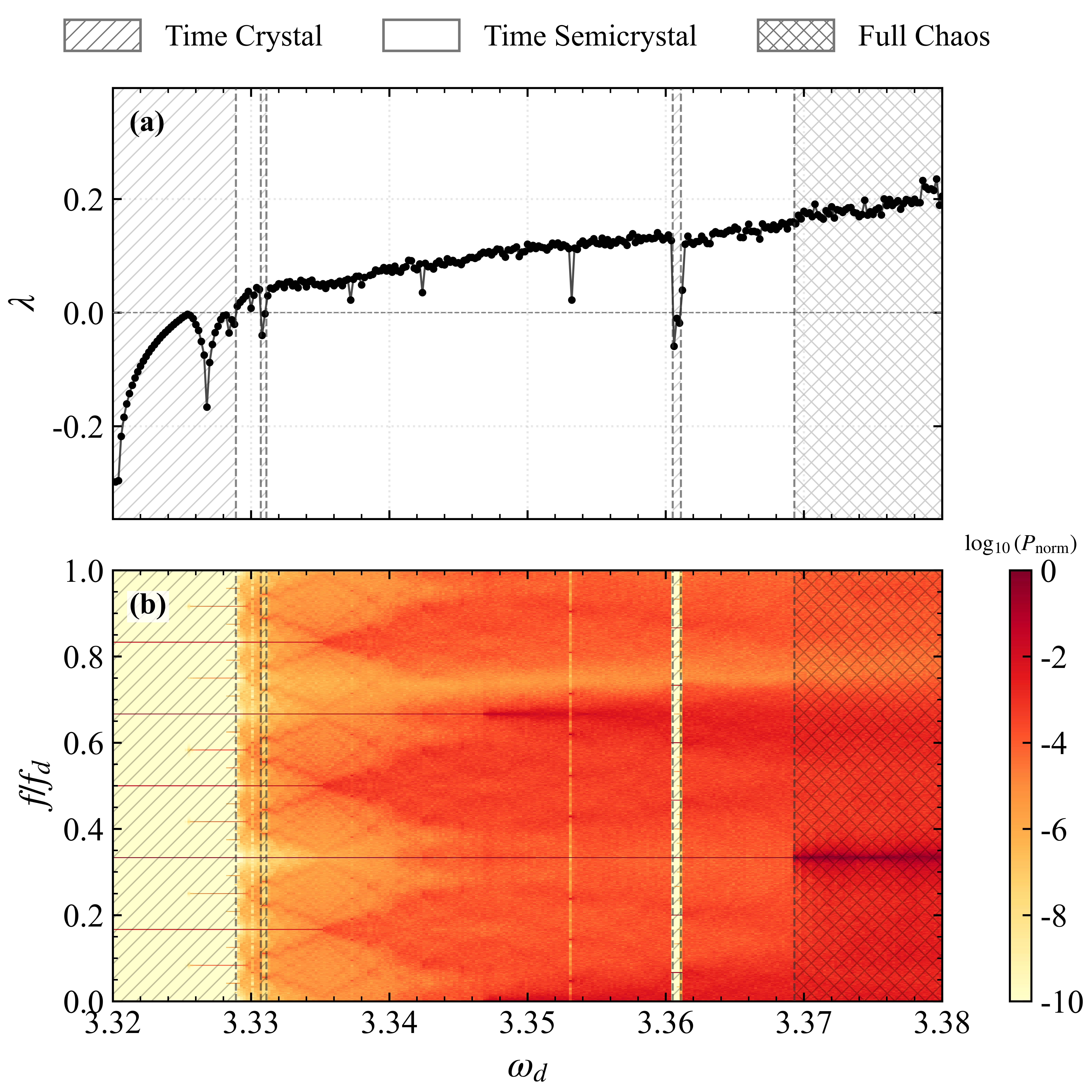}
    \caption{Lyapunov exponent and power spectrum as a function of $\omega_d$. Shaded regions delineate the dynamical regimes. (a) Largest Lyapunov exponent $\lambda$. (b) Density plot of $\log_{10} P_{\text{norm}}$ versus the normalized frequency $f/f_d$.}
    \label{fig:phase_diagram}
\end{figure}

Discarding transient dynamics, we analyze the steady-state evolution across varying driving angular frequencies $\omega_d$. As presented in \mpref{fig:phase_diagram}, the largest Lyapunov exponent (LLE) $\lambda$ and the normalized power spectral density $P_{\text{norm}}(f) = P(f)/P(f_d)$ collectively delineate three distinct dynamical regimes:

(i) \textbf{Discrete Time Crystal.} The system locks into stable periodic orbits ($\lambda < 0$), manifesting as a purely discrete subharmonic spectrum. Notably, distinct fractional peaks identify the emergence of higher-order (e.g., 6-period) time crystals.

(ii) \textbf{Time Semicrystal.} This phase is identified by a positive LLE ($\lambda > 0$) coexisting with a hybrid spectral structure. The subharmonic peaks within the chaotic continuum signify the presence of remnant periodic order.

(iii) \textbf{Full Chaos.} The sharp subharmonic peaks smear out and dissolve into a broadband continuum, leaving fully disordered dynamics ($\lambda > 0$).

As $\omega_d$ deviates from the time crystal regime, $\lambda$ becomes positive. The spectrum transforms into a mixed state where peaks coexist with a continuous background, signifying the melting of temporal order into the time semicrystal phase. Upon further tuning, the continuous component grows while the discrete subharmonic peak decays, indicating the evolution towards full chaos. Furthermore, the dispersion of the $2/3$ harmonic peaks observed in \mpref{fig:phase_diagram}(b) signals a restoration of half-period antisymmetry. This behavior corresponds to a dynamical phase transition akin to those in hysteretic Ginzburg-Landau dynamics~\cite{Chakrabarti:1998hpp,PhysRevE.75.026202,PhysRevLett.131.116701}, highlighting the internal structure of this remnant temporal order.

\begin{figure}[htp!]
    \centering
    \includegraphics[width=1.0\linewidth]{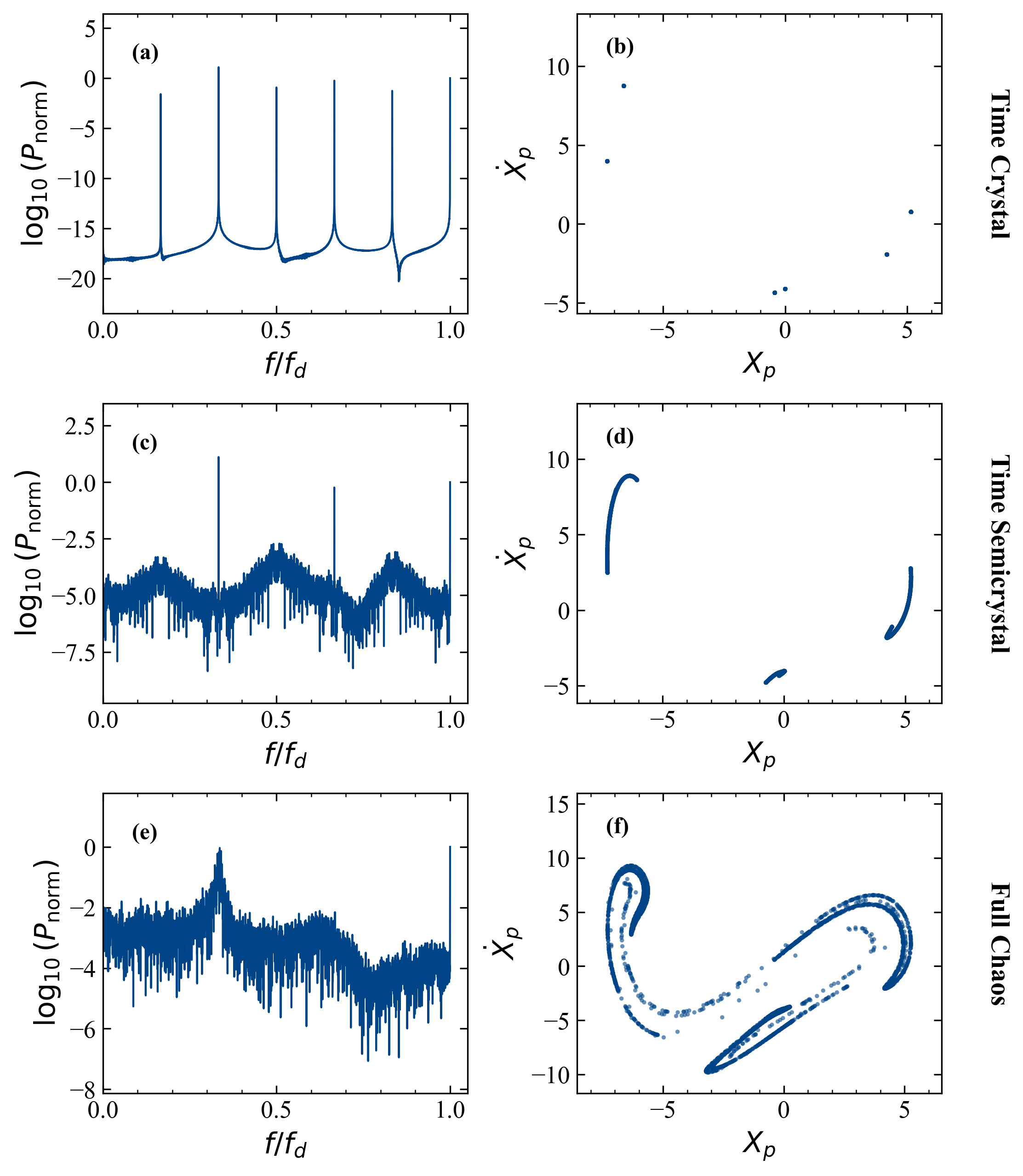}
    \caption{Normalized power spectra $\log_{10}(P_{\text{norm}})$ (left) and stroboscopic Poincar\'e sections $(X_p, \dot{X}_p)$ (right) characterizing the three dynamical phases. (a),(b) Discrete time crystal ($\omega_d = 3.324$). (c),(d) Time semicrystal ($\omega_d = 3.340$). (e),(f) Full chaos ($\omega_d = 3.378$).}
    \label{fig:signatures}
\end{figure}

We construct a stroboscopic map by sampling the boundary response at integer multiples of the driving period $T_d = 2\pi/\omega_d = 1/f_d$, yielding the discrete sequence
\be
X_p \equiv X(t_0 + pT_d), \quad p \in \mathbb{N},
\ee
where the initial time $t_0$ is chosen to align with the peak driving amplitude.
Based on the power spectra and stroboscopic Poincar\'e sections $(X_p, \dot{X}_p)$, we clearly demonstrate the characteristics of the time semicrystal. As shown in \mpref{fig:signatures}(a), the power spectral density of the discrete time crystal exhibits only discrete subharmonic peaks, whereas the full chaos phase in \mpref{fig:signatures}(e) is dominated by a continuous spectrum. The time semicrystal power spectrum in \mpref{fig:signatures}(c) displays a coexistence of discrete subharmonic peaks and continuous spectral components. Correspondingly, in the stroboscopic Poincar\'e sections, the finite isolated fixed points characterizing the DTC in \mpref{fig:signatures}(b) smear out into non-closed, band-like strange attractors in the time semicrystal regime, as shown in \mpref{fig:signatures}(d). These band-like structures encode the asymptotic periodicity of the surviving periodic skeleton within the chaotic background~\cite{PROVATAS1991295}. As the periodic skeleton completely dissolves, the system transitions into the chaotic fractal point cloud in \mpref{fig:signatures}(f).

These results jointly establish the time semicrystal as a distinct dynamical phase, whose essence is the coexistence of periodic skeleton and disorder. Therefore, clarifying the formation and stability mechanism of the time semicrystal phase provides key clues for understanding how temporal order degrades under driving and dissipation. It also lays the critical foundation for investigating the subsequent dynamical phase transitions and critical behaviors associated with this unique state.

\emph{Dynamical phase transition and scaling behavior}---In periodically driven holographic systems, the spontaneous emergence of nonequilibrium dynamical phases yields a rich phase diagram that includes discrete time crystal, time semicrystal and full chaos. A central question is whether the transitions between these phases can be quantitatively characterized, and whether they exhibit scaling laws analogous to equilibrium critical phenomena. The discussion focuses on representative dynamical transitions involving the melting of a DTC into a TSC, where subharmonic order yields to dynamical disorder, and the reorganization of periodic skeletons within the TSC phase, exemplified by the 6-TSC to 3-TSC transition. These transitions highlight the defining property of the time semicrystal phase, whose diagnostic feature is the coexistence of discrete subharmonic peaks with a continuous spectral background, reflecting the competition between residual periodic skeletons and disordered fluctuations.

\begin{figure}[htp!] 
    \begin{center}
    \includegraphics[width=1.0\linewidth]{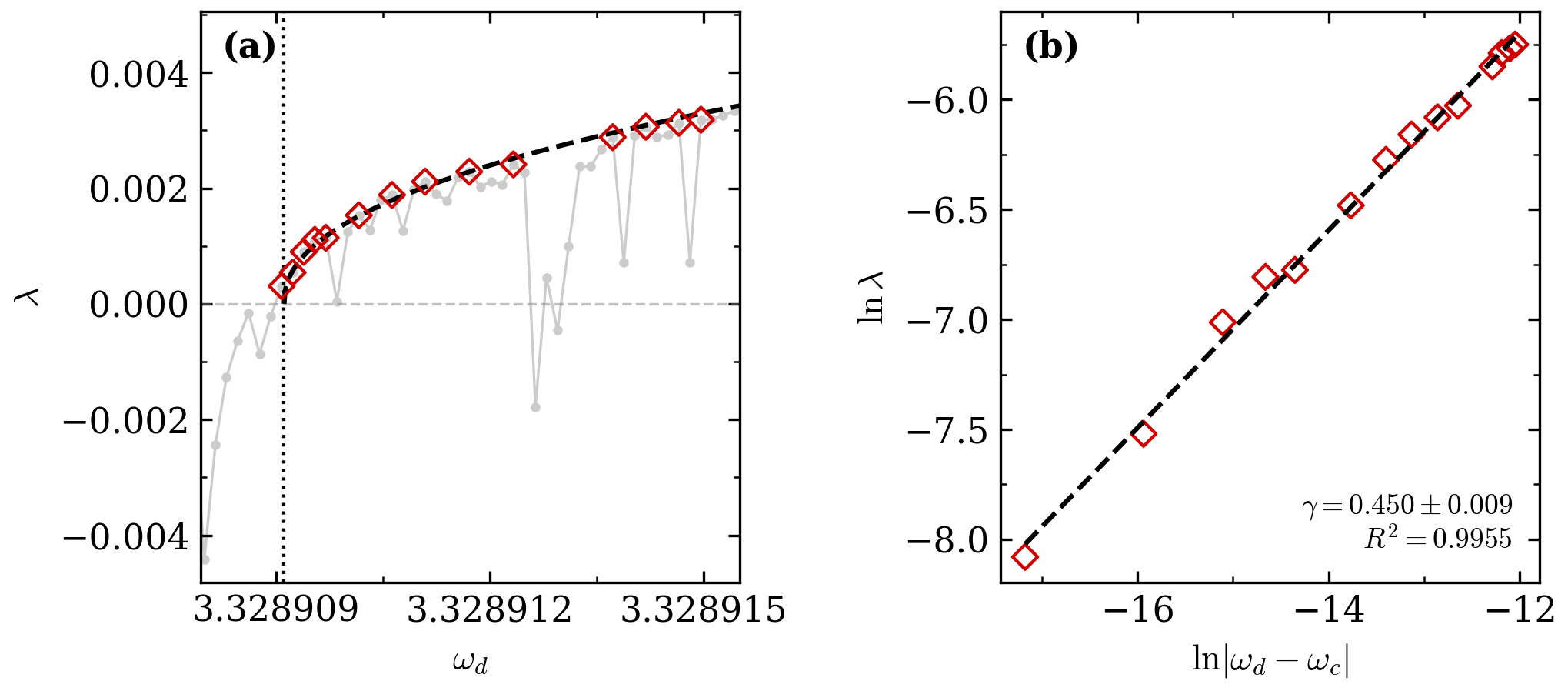}
    \end{center}
	\caption{Critical scaling of the DTC-TSC transition. (a) $\lambda$ vs $\omega_d$. The critical point $\omega_c$ is indicated by the vertical dotted line, and the envelope for $\lambda>0$ is shown by the red diamonds. (b) Power-law fit (dashed line) of the extracted envelope.}\label{fig:LEs}
\end{figure}

As $\omega_d$ is tuned, the DTC destabilizes via a cascade of period-doubling bifurcations, marking the entry into the TSC phase. This transition is characterized by a frequency spectrum evolving from discrete peaks into a mixed structure, alongside the emergence of temporal disorder. We employ the maximal Lyapunov exponent $\lambda$ as a dynamical order parameter to quantify this transition. As shown in \mpref{fig:LEs}(a), the raw $\lambda(\omega_d)$ near the critical point confirms the onset of disorder, punctuated by periodic windows~\cite{Huberman:1980dg}.
To extract the asymptotic scaling, we analyze the \textit{envelope} of the Lyapunov exponent, which filters out these local fluctuations and follows a power law, as shown in \mpref{fig:LEs}(b)
\be 
    \lambda \sim \lvert \omega_d - \omega_c \rvert^{\gamma},
\ee
where the critical parameter is $\omega_c \approx 3.329 $. The fitted critical exponent $\gamma = 0.450 \pm 0.009$ is in excellent agreement with the universal value for period-doubling cascades~\cite{Huberman:1980dg}.

This result indicates that the observed melting from a DTC into a TSC is not a simple crossover but a nonequilibrium dynamical phase transition governed by an explicit scaling law. Once $\omega_d$ crosses $\omega_c$, the system enters a mixed phase where temporal disorder coexists with residual periodic skeleton. Employing $\lambda$ as a dynamical order parameter reveals a critical power-law behavior strongly analogous to equilibrium critical phenomena.

\begin{figure}[htp!] 
    \begin{center}
    \includegraphics[width=1.0\linewidth]{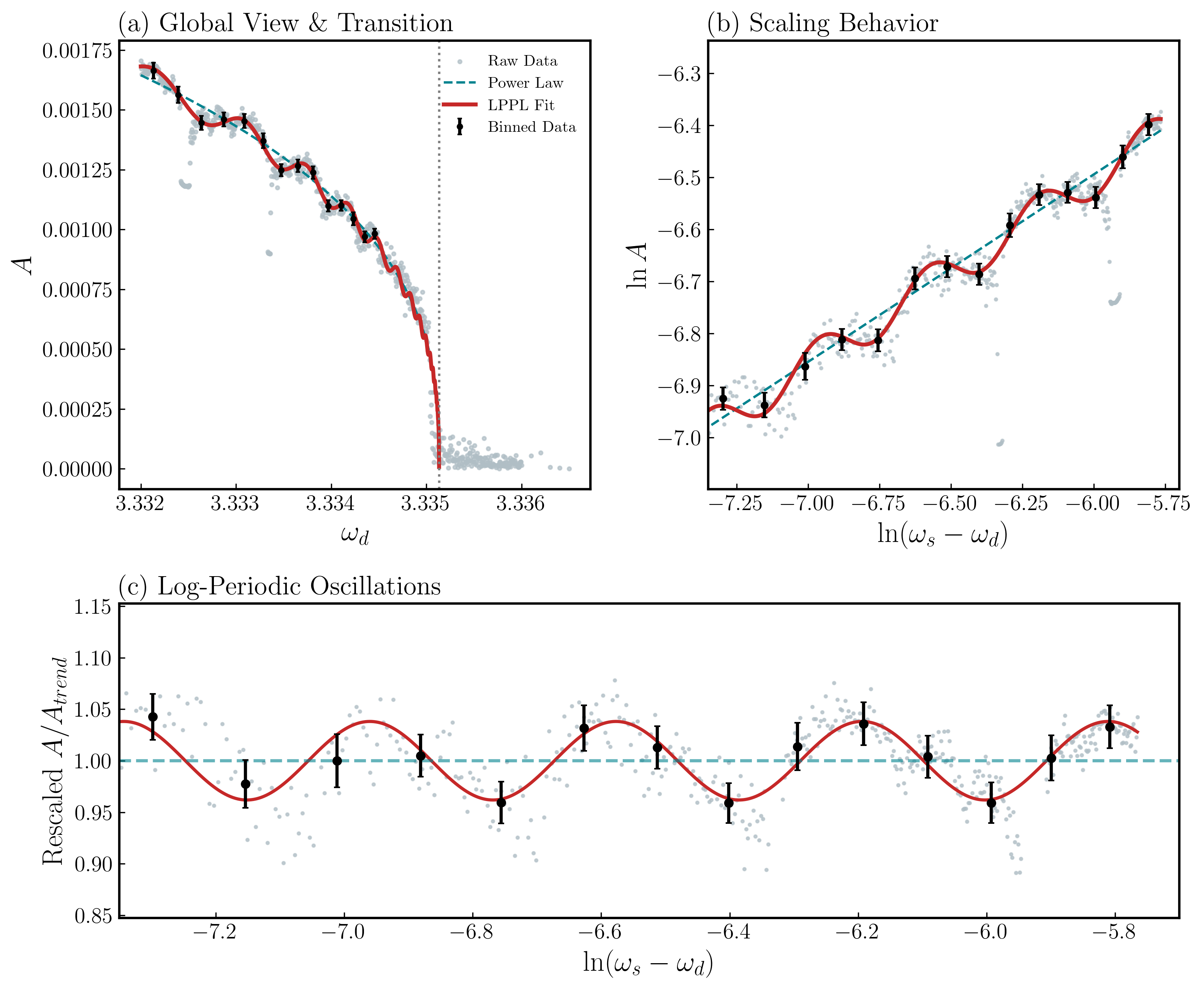}
    \end{center}
	\caption{Dynamical phase transition from 6-TSC to 3-TSC. (a) Order parameter $A$ vs $\omega_d$ for raw (gray dots) and binned (black circles) data. The critical point $\omega_s$ is indicated by the vertical dotted line. (b) $\ln A$ vs $\ln(\omega_s - \omega_d)$. (c) Rescaled amplitude $A/A_{\text{trend}}$ vs $\ln(\omega_s - \omega_d)$.}
\label{fig:TSCs PT}
\end{figure}

Far from being a featureless intermediate state, the time semicrystal phase manifests a structured transition. In this regime, discrete subharmonic peaks persist atop a continuous spectrum, eventually melting and reorganizing as $\omega_d$ is tuned. The transition from 6-TSC to 3-TSC is considered in detail. To quantify this internal transition, we define an order parameter $A(\omega_d)$ as the normalized spectral weight in a narrow band around the $1/6$ subharmonic
\be
A\equiv
\frac{\int_{f_d/6-\Delta f}^{f_d/6+\Delta f} P_{\text{norm}}(f)\,df}
{\int_{0}^{f_d} P_{\text{norm}}(f)\,df},
\ee
where the integrals are evaluated as a discrete sum over a few neighboring Fourier bins set by the finite time window. In \mpref{fig:TSCs PT}(a), gray points denote the raw data and black points with error bars denote the results after data binning, while a vertical dotted line marks the fitted critical point $\omega_s$. To mitigate numerical fluctuations arising from finite-time data, we employ a binning scheme on the logarithmic scale. Extracting the median and the standard error derived from the median absolute deviation for each bin suppresses outliers and clarifies the asymptotic critical trend.

After binning, the order parameter first follows a dominant power law trend, described by the dashed line in \mpref{fig:TSCs PT}(a)(b),
\be 
A_{\mathrm{trend}} = A_0(\omega_s-\omega_d)^{\nu}.\label{eq:Ftrend}
\ee 
The deviations from the leading power-law behavior are systematic rather than random. Normalized by the power-law trend, the residuals exhibit pronounced log-periodic oscillations against $\ln(\omega_s-\omega_d)$, as in \mpref{fig:TSCs PT}(c). Such oscillations are a hallmark of discrete scale invariance, a significant feature that has attracted considerable interest for its role in characterizing the fine structure of various critical transitions~\cite{Karevski_1996,Fernandes:2011qvq,Ammon:2018wzb}. Consequently, we fit the data using a critical scaling ansatz with log-periodic corrections
\be 
A = A_{\mathrm{trend}}\left[1+\alpha\cos\left(\Omega \ln(\omega_s-\omega_d)+\theta_0\right)\right].
\label{eq:lppl}
\ee
By treating $A_0$, $\alpha$, and $\theta_0$ as auxiliary parameters in the fit to \meqref{eq:Ftrend} and (\ref{eq:lppl}), we obtain the key parameters governing the underlying scaling behavior
\be 
\begin{aligned}
\omega_s &= 3.335134 \pm 0.000028, \quad &\nu &= 0.360 \pm 0.017, \\
\Omega &= 16.40 \pm 0.71. &
\end{aligned}
\ee
The critical exponent $\nu$ characterizes the underlying scaling behavior of the transition, while the log-periodic frequency $\Omega$ captures the discrete scaling. We can extract the discrete scaling factor $\delta_{\text{eff}} = \exp(2\pi/\Omega) \approx 1.47$, which characterizes the fundamental scale transformation in this dynamical transition. This characteristic factor is jointly determined by the system's near-critical properties and the chaotic attractors~\cite{Grebogi:1987,Krawiecki:2004}.

The observed melting of the periodic skeleton within the TSC phase is not a simple loss of order, but rather resembles a process of hierarchical reorganization. During this dynamical transition, the asymptotic decay of the specific periodic skeleton is captured by a leading power law, while log-periodic corrections provide a signature of its fine structure. The emergence of log-periodic oscillations in this nonequilibrium phase transition reflects the system's discrete scale invariance, which is equivalent to complex critical exponents in the framework of critical phenomena~\cite{SORNETTE1998239}. The melting of the $6$-period skeleton is governed by a discrete scaling behavior. This can be understood from the chaotic dynamics, where the intrinsic discrete scale invariance of the fractal sets of strange attractors translates directly into log-periodic modulations in the critical behavior of the order parameter~\cite{KRAWIECKI200389}. Near the critical point, the dynamics are governed by discrete scale invariance with a scaling factor $\delta_{\text{eff}}$, revealing the self-similar nature of the residual order in the time semicrystal phase.

\emph{Conclusion}---In a holographic system with spontaneous $Z_2$ symmetry breaking, a driven-dissipative setting gives rise to a rich dynamical phase diagram encompassing (higher-order) discrete time crystal, time semicrystal, and full chaos. By naturally incorporating dissipative effects via the black hole horizon, our holographic setup offers a rigorous framework to track the degradation of temporal order. Our results demonstrate that time semicrystal is not merely a crossover regime but a genuine dynamical phase with a clear physical meaning. The TSC phase with coexisting ``periodic skeleton + chaos'' structure emerges as DTCs melt, with the onset of disorder governed by power-law scaling. Within the time semicrystal, the transitions between distinct periodic skeletons follow scaling laws supplemented by log-periodic corrections, revealing a discrete scale invariance hierarchy. These results show that the time semicrystal is a dynamical phase with critical scaling behavior and internal hierarchical structure. This study offers a new avenue for investigating the degradation of temporal order in driven-dissipative systems and opens a route to a sharper understanding of orders in chaos.

Our findings therefore suggest that time semicrystal order and its hierarchical melting can be realized and tested in driven-dissipative quantum platforms. Establishing direct links between the diagnostics introduced here and experimental observables, and probing their robustness across broader dissipative and driving conditions, will be an essential step towards confronting holographic predictions with experiments.

\begin{acknowledgments}
We would like to thank SangJin Sin, Jianxin Lu, Minxin Huang, Surojit Dalui, Shuta Ishigaki, Jia-Geng Jiao and Peng Yang for helpful discussions. XHG was partially supported by the National Natural Science Foundation of China (NSFC) (Grant Nos.~12275166 and 12311540141). YT was partially supported by NSFC, China (Grant Nos.~12375058 and 12361141825). YQL was partially supported by NSFC, China (Grant No.~12405072) and China Postdoctoral Science Foundation (Grant No.~2024M761914).

\end{acknowledgments}


\newpage

\onecolumngrid
\appendix
\section*{SUPPLEMENTARY MATERIAL}
\setcounter{equation}{0}
\renewcommand{\theequation}{S\arabic{equation}}

This Supplementary Material details the explicit equations of motion (EOMs), our numerical discretization scheme, and the algorithm for evaluating the largest Lyapunov exponent.

Simulating the nonequilibrium dynamics requires solving the bulk EOM for the rescaled scalar field $\psi(t,z) = \phi/z$. Assuming spatial isotropy in Eddington-Finkelstein coordinates, the dynamics are governed by the nonlinear partial differential equation
\begin{equation}
2\partial_t\partial_z\psi - h(z)\,\partial_z^2\psi - h'(z)\,\partial_z\psi + z\psi + \frac{1}{3}\psi^3 = 0,
\label{eq:sm_eom}
\end{equation}
with $h(z)=1-z^3$ being the blackening factor. At the AdS boundary ($z=0$), the interplay between the double-trace deformation and the external periodic drive $F_d(t) = F_0 \sin(\omega_d t)$ enforces the time-dependent mixed boundary condition
\begin{equation}
\left.\partial_t\psi\right|_{z=0} = \left.(\partial_z\psi - \bar{\varkappa}\psi + F_d(t))\right|_{z=0}.
\label{eq:sm_bc}
\end{equation}
Throughout our simulations, we fix the deformation parameter at $\bar{\varkappa}=-2$ and the drive amplitude at $F_0=4$.

For the spatial discretization along the holographic coordinate $z \in [0, 1]$, we utilize a Chebyshev pseudo-spectral method. The physical domain is mapped onto the standard interval $[-1, 1]$, and the field is discretized across $N=25$ collocation points, allowing the spatial derivatives $\partial_z$ and $\partial_z^2$ to be evaluated via exact differentiation matrices. Time integration is performed using a standard fourth-order Runge-Kutta (RK4) scheme with a fixed time step $\Delta t =T_d/8000$, where $T_d =2\pi / \omega_d$ is the driving period. To ensure the system has relaxed to a nonequilibrium steady state, we strictly discard the early-time transient evolution prior to extracting the boundary observable $X(t) = \psi(t, z=0)$ at a sampling interval of $T_d/80$.

To characterize temporal disorder in the time semicrystal and fully disordered phases, we calculate the largest Lyapunov exponent (LLE) $\lambda$. This is achieved by combining linear perturbation evolution with the Benettin algorithm. Linearizing the system around the base trajectory $\psi(t, z)$ yields the variational equation for a small perturbation $\delta\psi(t, z)$
\begin{equation}
    2\partial_t \partial_z(\delta\psi) - h \partial_z^2(\delta\psi) - h' \partial_z(\delta\psi) + z(\delta\psi) + \psi^2\delta\psi = 0.
\end{equation}

The state-independent external drive $F_d(t)$ vanishes from the linearized perturbation equation. This leaves a homogeneous boundary condition at $z = 0$
\begin{equation}
    \partial_t(\delta\psi)|_{z=0} = \left(\partial_z(\delta\psi) - \bar{\varkappa}\delta\psi\right)|_{z=0}.
\end{equation}

The LLE $\lambda$ characterizes the average exponential growth rate of this perturbation in the tangent space. Its theoretical definition is
\begin{equation}
    \lambda = \lim_{t \to \infty} \frac{1}{t} \ln \frac{\|\delta\psi(t)\|}{\|\delta\psi(0)\|}, \label{Ly_def}
\end{equation}
where $\|\cdot\|$ denotes the norm of the finite-dimensional discretized field, defined as 
$\|\delta\psi\| = \sqrt{\sum_{i=1}^{N} |\delta\psi(z_i)|^2}$ with $N=25$ collocation points. Direct long-time integration of the variational equation leads to numerical overflow in chaotic regions. We avoid this through standard periodic renormalization. The initial perturbation $\delta \psi(0)$ is initialized as a Gaussian random perturbation with amplitude scaled to the initial norm $\epsilon = 10^{-8}$. The system is then evolved over an initial transient window. This step ensures the perturbation vector naturally aligns with the most unstable direction in the tangent space prior to data collection. After this transient phase, we accumulate the growth rate over a sufficiently large number of driving periods. At the end of each driving period, the perturbation is immediately renormalized back to the initial norm $\epsilon$. The time average of these accumulated logarithmic growths converges to the theoretical value in Eq.~\ref{Ly_def}. The sign of $\lambda$ provides a quantitative diagnostic for temporal disorder. A strictly positive $\lambda$ signifies sensitivity to initial conditions. This diagnostic distinguishes the regular time crystal phase discussed in the main text from the temporally disordered time semicrystal and fully chaotic phases.

\end{document}